\documentclass[aps,prl,twocolumn,groupedaddress,showpacs]{revtex4}
\usepackage{graphicx}

\begin{document}

\title{Pressure-induced $s$-band ferromagnetism in alkali metals}

\author{Chris J. Pickard}

\affiliation {Department of Physics and Astronomy, University College
London, Gower St, London WC1E 6BT, United Kingdom}

\author{R. J. Needs}

\affiliation {Theory of Condensed Matter Group, Cavendish Laboratory,
J J Thomson Avenue, Cambridge CB3 0HE, United Kingdom}

\date{\today}

\begin{abstract}
  First-principles density-functional-theory calculations show that
  compression of alkali metals stabilizes open structures with
  localized interstitial electrons which may exhibit a Stoner-type
  instability towards ferromagnetism.  We find ferromagnetic phases of
  the lithium-IV-type, simple cubic, and simple hexagonal structures
  in the heavier alkali metals, which may be described as $s$-band
  ferromagnets.  We predict that the most stable phases of potassium
  at low temperatures and pressures around 20 GPa are ferromagnets.
\end{abstract}


\maketitle

Among the elements, bulk ferromagnetism is found only in the first row
of the transition metals and in the lanthanides, where partially
filled $3d$ and $4f$ electronic shells of strongly-localized orbitals
are present.  Magnetic order is normally reduced and eventually
destroyed by the application of pressure because it tends to
delocalize electronic states.  The alkali metals are archetypal
nearly-free-electron materials in which each atom contributes a single
valence electron to a gas whose effective interaction with the ionic
cores is weak.  They might therefore be thought of as the least likely
elements in which to find bulk ferromagnetism.  The alkali metals,
lithium (Li), sodium (Na), potassium (K), rubidium (Rb), cesium (Cs),
and probably francium (Fr) (the most unstable naturally occurring
element) adopt body-centered-cubic (bcc) phases under ambient
conditions which, on compression, transform to face-centered-cubic
(fcc) phases \cite{McMahonN2006_review}.  Diamond-anvil-cell
experiments have shown, however, that they adopt more open structures
at higher pressures \cite{McMahonN2006_review}.  Density-functional
theory (DFT) calculations have reproduced the stability of the
experimentally observed phases and have given insights into their
electronic structures
\cite{NeatonA99,Neaton_2001,HanflandSCN00,PickardN09_lithium,Ma_2009}.
The most surprising result is that a large amount of valence charge in
the open structures resides within the interstitial regions rather
than in close proximity to the ions.  This corresponds to the
formation of ``electrides'' in which the interstitial electrons form
the anions
\cite{Von_Schnering_1987,RousseauA08,PickardN09_lithium,Ma_2009}.  The
open structures correspond to well-packed ionic solids when both the
alkali metal ions and the centers of the interstitial electronic
charges are designated as ionic positions \cite{PickardN10_aluminium}.

The band structures of alkali metals deviate substantially from
nearly-free-electron behavior under applied pressure
\cite{NeatonA99,Neaton_2001,HanflandSCN00,PickardN09_lithium,Ma_2009}.
The occupied valence bands become flatter than the corresponding
nearly-free-electron ones, which narrows the occupied valence
bandwidth.  Ashcroft and coworkers
\cite{NeatonA99,Neaton_2001,RousseauA08} have attributed this
phenomenon to the interaction of the valence electrons with the
relatively incompressible ionic cores that occupy an increasingly
large fraction of the total volume as pressure is increased.  Under
ambient conditions the effective interaction between the valence
electrons and ionic cores is weak, but it becomes strongly repulsive
under pressure and forces valence electrons to occupy interstitial
positions \cite{NeatonA99,Neaton_2001}.  The interstitial regions in
close-packed structures are numerous, but small, and ``cutting up''
the valence charge into small regions increases the kinetic energy.
The kinetic energy can, however, be reduced by adopting more open
structures which have less numerous but larger interstitial regions in
which to accommodate the valence electrons.  This effect evidently
overcomes the concomitant increase in core-core repulsion energy.

The Fermi surface/Bragg plane mechanism for stabilising structures
\cite{Jones_1934} involves the action of the Fourier components of the
lattice potential on the degenerate electron orbitals on the Bragg
planes in reciprocal space.  The total energy is lowered if a
structure is adopted in which Bragg planes graze the Fermi surface,
because the occupied orbitals just inside the Bragg plane are lowered
in energy while those of the unoccupied orbitals just outside are
raised.  This mechanism can lead to structural instabilities and it is
likely to be involved in determining the details of the high-pressure
structures of many of the alkalis \cite{Ackland_2004,Degtyareva_2009}.
The Fermi surface/Bragg plane mechanism normally operates when $g(E)$
is large in the region around the Fermi energy $E_{\rm F}$, but an
alternative instability comes into play when $g(E)$ is also strongly
peaked around $E_{\rm F}$.  In this case the energy may be lowered by
transferring electrons from one spin channel to the other, resulting
in ferromagnetism (FM).  This is the Stoner instability
\cite{Stoner_1939}.  The key quantities in Stoner's theory are
$g(E_{\rm F})$ and the effective interaction $I$ between the up and
down spin densities in the unpolarized state, with the instability to
FM occurring when $g(E_{\rm F}) I > 1$.

Motivated by the above analysis, we have investigated magnetic
ordering in the alkalis under high-pressures.  We searched for
low-enthalpy structures using \textit{ab initio} random structure
searching (AIRSS) \cite{PickardN06_silane,Airss_review}, which has
been successfully applied to systems as diverse as metals under high
pressures \cite{PickardN09_lithium,PickardN10_aluminium} and molecular
solids \cite{PickardN07_hydrogen,PickardN08_ammonia}.  We used the
\textsc{castep} plane-wave DFT code \cite{ClarkSPHPRP05} and the
Perdew-Burke-Ernzerhof (PBE) Generalized Gradient Approximation (GGA)
density functional \cite{PerdewBE96}, and we also present results
obtained with a Thomas-Fermi screened exchange functional (SX)
\cite{clark_2010} and the local spin density approximation (LSDA).  We
used ultrasoft pseudopotentials \cite{Vanderbilt90}, treating all
three electrons explicitly for Li, and nine electrons for the other
alkalis.  Brillouin zone integration grids of spacing
$2\pi\times$0.05~\AA$^{-1}$ were used for the searches and the
low-enthalpy structures were further relaxed using a finer grid
spacing of $2\pi\times$0.03~\AA$^{-1}$ which gave very accurate
enthalpy differences between the phases.  Large plane wave basis set
energy cutoffs were used.  The phonon calculations were performed
using a finite-displacement method and 64-atom supercells.

We performed spin-polarized calculations, starting some relaxations in
a high spin state with an average spin density of one electron per
atom and others with zero average spin density.  In each calculation
the spin density was allowed to evolve freely as the structure was
relaxed.  We also performed calculations without spin polarization.
The use of a wide variety of starting structures and spin states was
found to be important in allowing many different spin and atomic
configurations to be accessed.  The main searches were performed at
pressures around those at which experiments show that the fcc
structures transform to more open phases.  We performed searches with
unit cells containing up to eight atoms, relaxing a total of about
1500 structures.  Calculations for other known structures of the
alkalis which have more than eight atoms per cell were also performed,
namely a periodic structure of space group $I4/mcm$ with 56 atoms
which is a good analogue of the K-III (Rb-IV) incommensurate
host-guest structure, and the 84-atom $C222_1$ structure of Cs-III and
the related 52-atom structure of Rb-III \cite{McMahonN2006_review}.

We find good agreement with the experimentally observed phase
transitions.  Our calculated coexistence pressure for K-fcc and
K-$I4/mcm$ of 19.8 GPa is close to the experimental transition
pressure of 23 GPa \cite{McMahonN2006_review}.  We note that Marqu\'es
\textit{et al.}  \cite{Marques_2009} have reported finding the
K-$I4/mcm$ phase in some high-pressure experiments and the K-hP4 phase
in others, which is consistent with our finding that K-hP4 is only
slightly less stable than K-$I4/mcm$.  Our coexistence pressure for
the Rb-fcc and Rb-$C222_1$ phases of 14.2 GPa is in excellent
agreement with the experimental transition pressure of 14 GPa
\cite{McMahonN2006_review}.  We obtain a coexistence pressure for
Cs-fcc and Cs-$C222_1$ of 4.8 GPa, compared with the experimental
transition pressure of 4.2 GPa \cite{McMahonN2006_review}.  We find
Cs-IV to be stable from 4.8--10 GPa, in excellent agreement with
experiment \cite{McMahonN2006_review}.  We find no region of stability
for the observed Cs-III structure which in our calculations only
becomes more stable than the fcc phase at 4.9 GPa, although this is a
small discrepancy.  In general we find that our calculated coexistence
pressures can be brought into agreement with the experimental data by
rigidly shifting the enthalpy curves by less than 5 meV, which
indicates the high level of accuracy of our calculations.

\begin{figure}
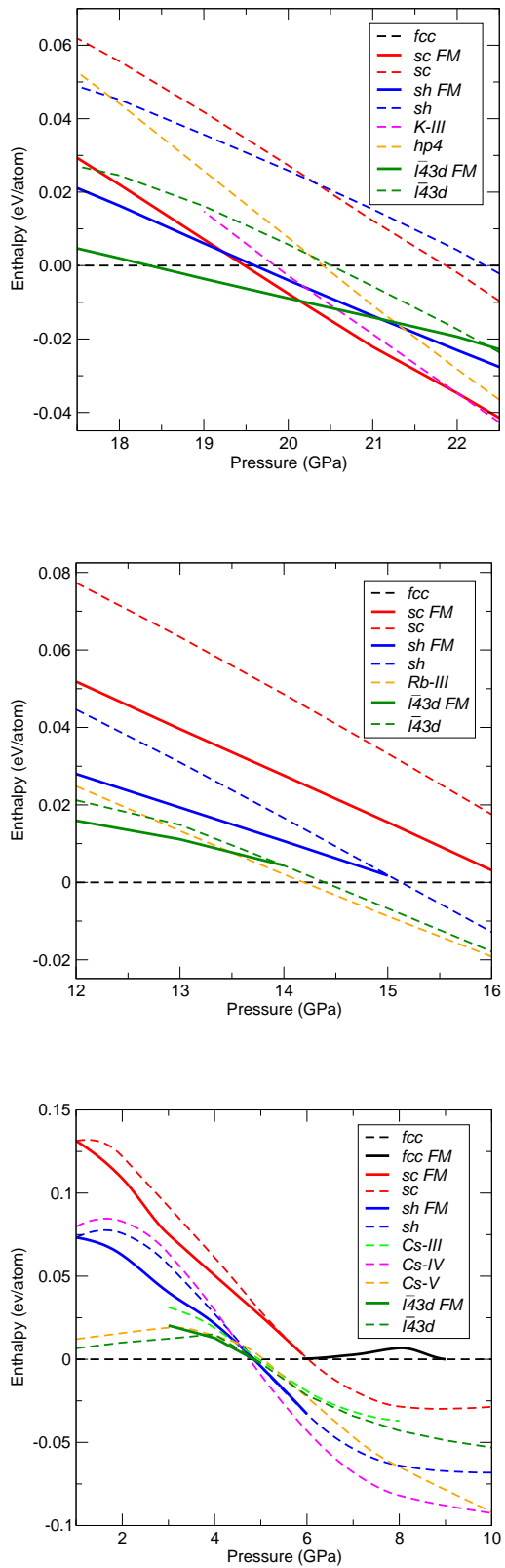

\includegraphics[width=0.375\textwidth]{Fig1a.eps}\\
\includegraphics[width=0.375\textwidth]{Fig1b.eps}\\
\includegraphics[width=0.375\textwidth]{Fig1c.eps}
\caption{\label{fig:enthalpy} (color online).  Enthalpy-pressure curves for
the FM and paramagnetic phases of K, Rb, and Cs.
Differences in enthalpy from the fcc phase are plotted.  FM phases are shown
as solid lines and paramagnetic phases as dashed lines.}
\end{figure}

We found $I\bar{4}3d$, simple cubic (sc) and simple hexagonal (sh)
phases of K, Rb, Cs and Fr with strong FM ordering, and a weakly FM
Cs-fcc phase, see Fig.\ \ref{fig:enthalpy}.  The enthalpy reductions
due to the formation of FM moments in the sc phases are similar to
those in the corresponding sh phases.  The sh phases are more stable
than the sc phases at lower pressures in K and Rb, and more stable at
all pressures in Cs.  We did not find any spin-polarized states of Li
or Na in our fully-converged calculations, but we can obtain them by
reducing the number of k-points.  This shows that Li and Na are close
to a FM instability and demonstrates the importance of carefully
studying the convergence with respect to the k-point sampling, as we
have done.  We predict that the FM K-$I\bar{4}3d$ phase to be the most
stable in the range 18.5--20 GPa and the FM K-sc phase to be the most
stable in the range 20--22 GPa.  The FM ordering leads to an energy
gain of a few tens of meV per atom.  When the centers of the
interstitial electronic charges are designated as ionic positions, sc
becomes the CsCl structure, sh the MgB$_2$ structure, and $I\bar{4}3d$
the Th$_3$P$_4$ structure \cite{Meisel_1939}, with the Cs, Mg and P
sites being those of the cations and the Cl, B and Th sites being
those of the interstitial electrons.

To explore the sensitivity to the density functional we also performed
calculations using a Thomas-Fermi SX functional \cite{clark_2010} and
the LSDA.  For the SX calculations we used a screening wave vector of
$k_s$ = 0.764 a.u., which corresponds to a Wigner-Seitz radius of
$r_s$ = 4.2 a.u.  This is larger than the value of $r_s$ = 3.4 a.u.\
obtained from the average valence charge density of FM K-sc at 20 GPa,
but the results are rather insensitive to reasonable variations in
$k_s$ as it is proportional to the one-sixth power of the average
charge density.  Using the SX functional instead of PBE-GGA stabilizes
FM K-sc over K-fcc by about 15 meV per atom.  We expect that the sc
phases of each of the alkalis would be further stabilized with respect
to the fcc phases when calculated with the SX functional.  We did not
find magnetic ordering in K when using the LSDA functional which
predicts the K-sc phase to be stable above about 20 GPa, in
disagreement with experiment.  Given that the pressures of the
observed transitions calculated within PBE-GGA are in very good
agreement with experiment, we are inclined to believe that it gives a
better description of compressed alkalis than the LSDA or SX
functionals.

Calculations of the harmonic vibrational modes of the K-fcc, K-sc and
FM K-sc phases showed them to be dynamically stable.  The phonon modes
of the sc phases are substantially softer than those of K-fcc, and
including the zero-point enthalpy stabilizes the K-sc and FM K-sc
phases over K-fcc by about 25 meV per atom at 20 GPa.  The stability
of K-sc and FM K-sc over K-fcc is slightly increased by including the
vibrational contribution to the free energy at a temperature of 300 K,
but the magnetic moment will be reduced by disordering of the spins.

\begin{figure}
\includegraphics[width=0.4\textwidth]{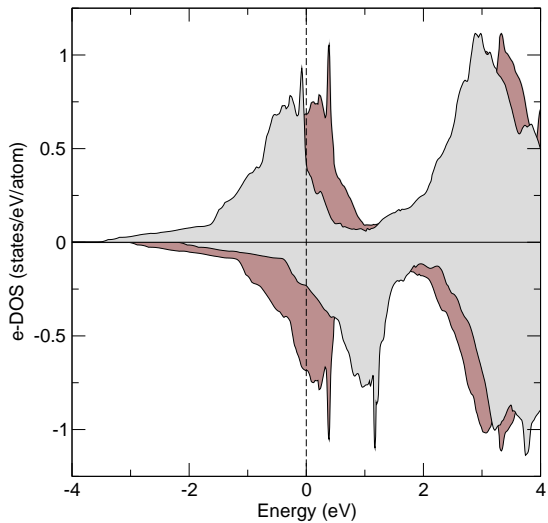}
\caption{\label{fig:eDoS_sc_K} (color online). The electronic density of
  states, $g(E)$, of K-sc at 20 GPa in states per eV per atom.  Data for the
  paramagnetic phase is shown in brown and for FM K-sc in grey.  The Fermi
  energy is shown as a vertical dotted line.}
\end{figure}

The valence electron density of states of the paramagnetic and FM
states of K-sc at 20 GPa are shown in Fig.\ \ref{fig:eDoS_sc_K}.  In
the paramagnetic system, $E_{\rm F}$ is close to the top of a large
peak in $g(E)$ and the system is ripe for a Stoner instability.  In
the FM K-sc phase $E_{\rm F}$ falls just above the peak for the
majority-spin band and well below the peak for the minority spin band.
The FM K-sc state is about 35 meV per atom more stable than the
paramagnetic K-sc phase, which is sufficient to make FM K-sc the most
stable phase in the pressure range 20--22 GPa.  The value of $g(E_{\rm
  F})$ for non-spin-polarized K-sc at 20 GPa is 1.4 per eV per atom,
so that the Stoner criterion is satisfied for $I > 0.7$ eV, which is
similar to the values deduced for FM in transition metals
\cite{Andersen_1977}.

\begin{figure}
\includegraphics[width=0.35\textwidth]{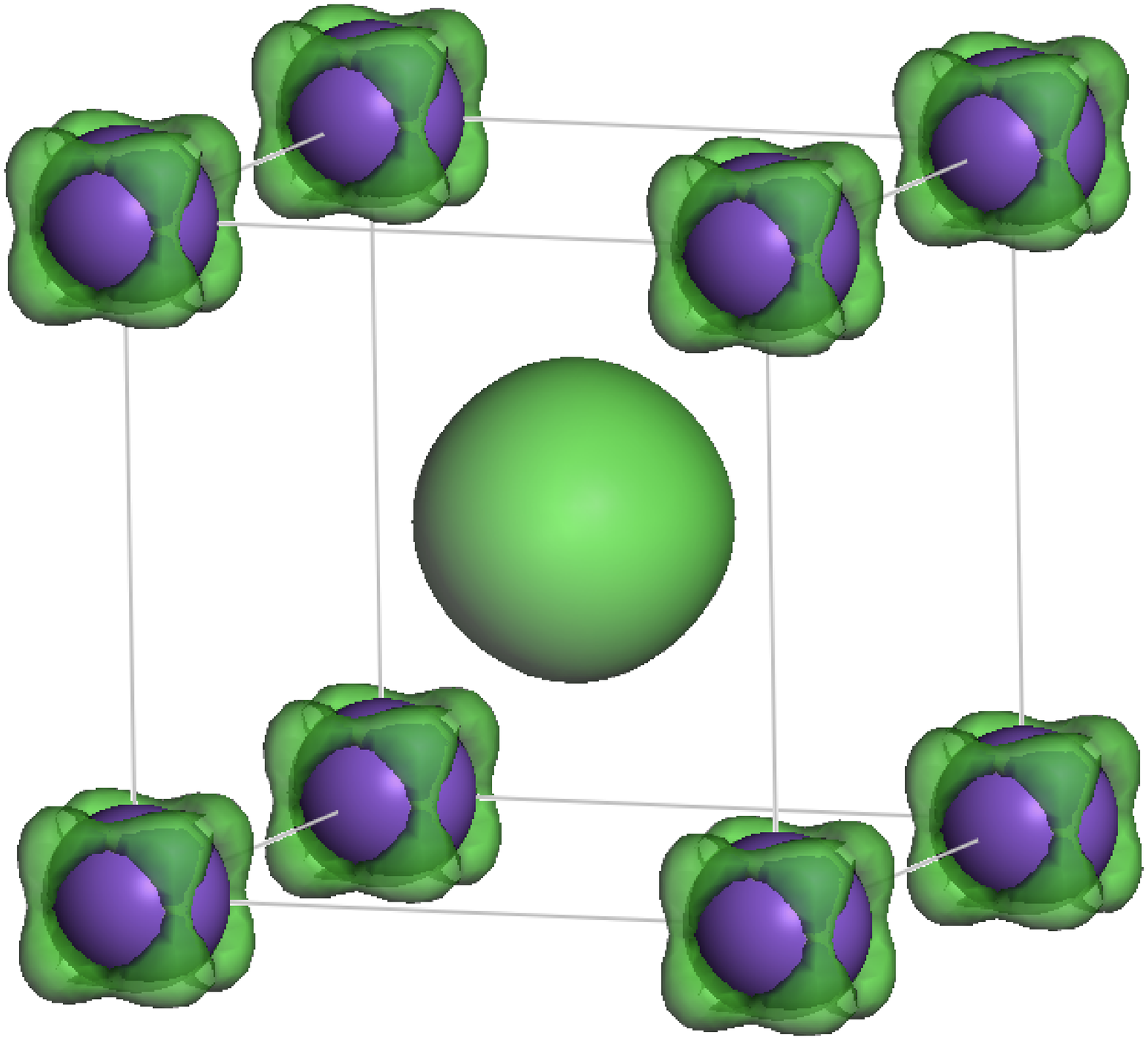}\\
\includegraphics[width=0.4\textwidth]{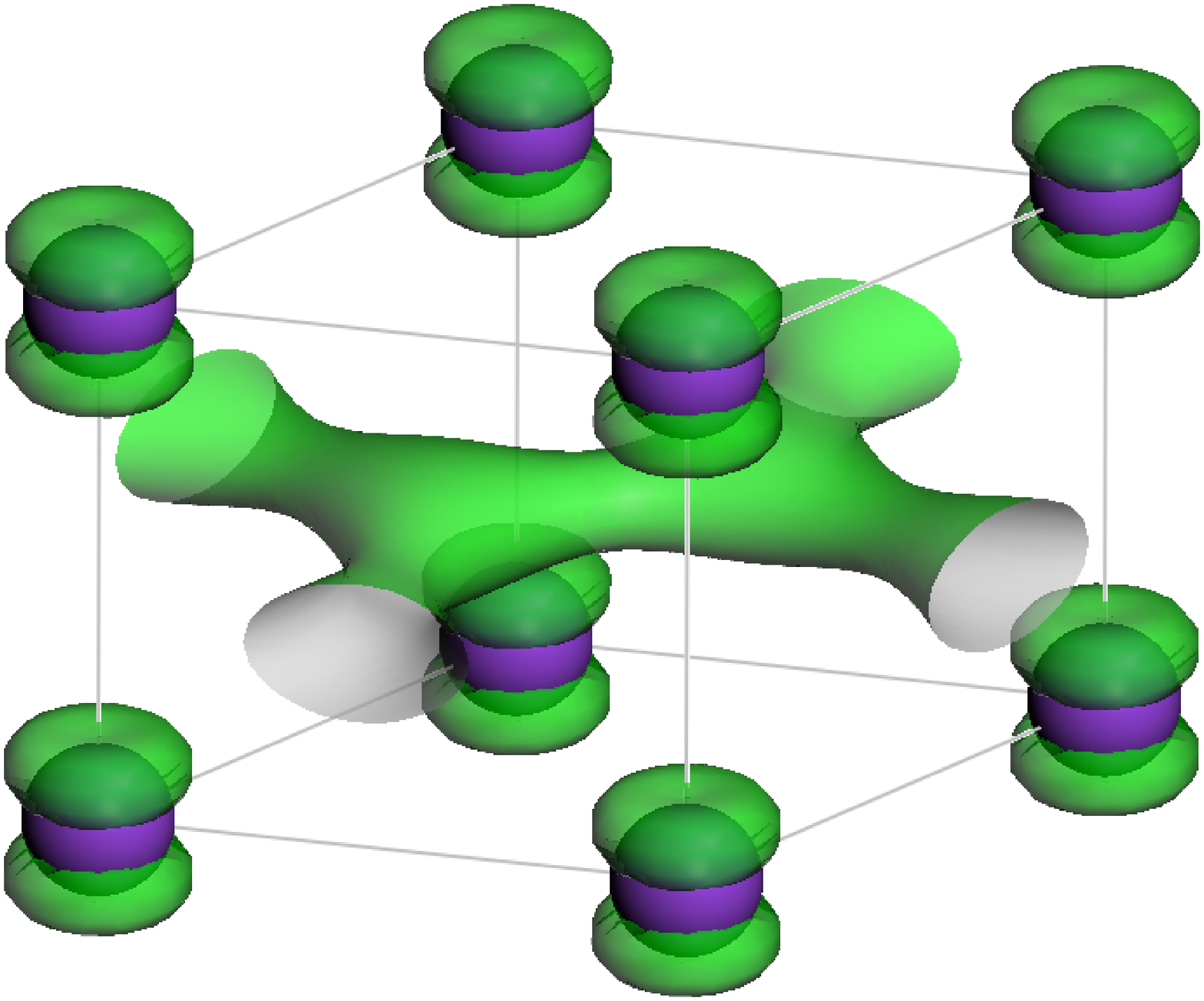}\\
\caption{\label{fig:spin_density} (color online). The spin density of FM K-sc
  (top) and FM K-sh (bottom) at 20 GPa.  The purple spheres represent the
  atoms and the spin density is shown in green.}
\end{figure}

The spin densities of FM K-sc and FM K-sh at 20 GPa are shown in Fig.\
\ref{fig:spin_density}.  A large blob of spin-polarized charge density
resides on the interstitial site at the center of the cube of K-sc,
while the spin polarization on the atomic sites is small.  The spin
polarization on the atoms is also small in FM K-sh, and the
interstitial spin density is more diffuse.  The FM K-sc structure
achieves its maximum spin moment of about 0.72 electrons per atom at
22 GPa, while the maximum spin moment in FM K-sh of about 0.62
electrons per atom occurs at 20 GPa, and the maximum spin moment of
K-$I\bar{4}3d$ is about 0.4 electrons per atom.
The FM phases might be described by a Hubbard-like model \cite{Hubbard_1963}
using tight-binding $s$ orbitals centered on the interstitial regions.  The
magnetic state would then be described as $s$-band FM, which is very different
from the $d$ or $f$ band FM observed in the transition metal and lanthanide
elements.

We are not aware of any experimental evidence for magnetic ordering in
bulk alkali metals, although there is evidence of it in
low-dimensional systems.  Theoretical work by Overhauser
\cite{Overhauser_1971} suggested that charge-density and
spin-density-wave instabilities might occur in alkali metals at low
pressures, but these ideas have not been widely accepted.  A DFT study
by Zabala \textit{et al}.\ \cite{Zabala_1998} found spontaneous
magnetization in ``Na wires'' modelled by jellium, and Bergara
\textit{et al}.\ \cite{Bergara_2003} found instabilities to FM in
atomically-thin Li and Na wires.  FM ground states are, however,
forbidden in strictly one-dimensional systems by the Lieb-Mattis
theorem \cite{Lieb-Mattis_1962} and, in quasi-one-dimensional systems,
quantum fluctuations tend to suppress FM.  Experimental observations
of FM in K clusters incorporated within a zeolite \cite{Nozue_1992}
and anti-FM in K clusters in a nanographite-based host
\cite{Takai_2007} have also been reported.  We have used the AIRSS
method to search for structures of small unsupported K clusters,
finding that some weakly anti-FM states are energetically favorable
and that quite strongly FM phases with moments up to about 0.5
electrons per atom occur at higher energies.  Electride formation in
alkali metal clusters at ambient pressure is highly unlikely because
the bonds are too long, and the magnetic ordering must produced by
some other mechanism.  It might be possible to further stabilize FM
phases of bulk alkali metals by alloying them with other alkalis or
other species.  Low-temperature experiments looking for magnetism in
compressed alkalis are needed to test our predictions.

In summary, we predict mechanically-stable FM phases of the heavier
alkali metals to have low enthalpies at pressures just above the
stability range of the fcc phases.  FM phases with the Li-IV
($I\bar{4}3d$), sc, and sh structures can be described as $s$-band
electride ferromagnets.  The FM K-$I\bar{4}3d$ and K-sc phases are
predicted to be the most stable at low temperatures and pressures
around 20 GPa.

\begin{acknowledgments}
  The authors were supported by the Engineering and Physical Sciences
  Research Council (EPSRC) of the UK.
\end{acknowledgments}

\end{document}